\documentclass{article}

\usepackage[eandd, final]{neurips_2026}

\usepackage[utf8]{inputenc}
\usepackage[T1]{fontenc}
\usepackage{hyperref}
\usepackage{url}
\usepackage{booktabs}
\usepackage{amsmath}
\usepackage{amssymb}
\usepackage{amsfonts}
\usepackage{nicefrac}
\usepackage{graphicx}
\usepackage{subcaption}
\usepackage{xcolor}
\usepackage{multirow}
\usepackage{microtype}

\newcommand{\eg}{\textit{e.g.}}

\newcommand{\Kset}{\{128,256,512,1024\}}
\newcommand{\MSVQVAE}{MS-VQ-VAE}

\definecolor{emphblue}{RGB}{33,102,172}

\title{Codebook Capacity Governs Perceptual Quality Across Resolutions\\
       in Hierarchical Discrete Video Compression}

\author{%
  Manikanta Kotthapalli \\
  Department of Computer Science\\
  Portland State University\\
  Portland, OR 97201 \\
  \And
  Banafsheh Rekabdar \\
  Department of Computer Science\\
  Portland State University\\
  Portland, OR 97201 \\
}

\begin{document}
\maketitle

% ─────────────────────────────────────────────────────────────────────────────
\begin{abstract}
Learned video codecs based on continuous latent representations typically
require resolution-specific retraining or rate--distortion (RD) recalibration
when scaling to new spatial resolutions, because entropy models and Lagrangian
weights are tightly coupled to the operating point.
We investigate whether hierarchical \emph{discrete} latent codecs exhibit
the same sensitivity.
Using a controlled empirical study of \MSVQVAE{} video compression across
codebook sizes $K \in \Kset$ and resolutions $64{\times}64$, $128{\times}128$,
and $256{\times}256$ on UCF101, we show that perceptual quality (LPIPS)
depends strongly on codebook capacity but only negligibly on spatial resolution.
Fitting a log-linear model $Q(K,r) = \alpha\log_2\!K + \beta\log_2\!r + \gamma$
to all 12 operating points yields $\alpha{=}{-}0.0094$ ($t{=}{-}6.6$, $p{<}0.001$)
and $\beta{=}{-}0.0009$ ($t{=}{-}0.43$, $p{=}0.68$, not significant),
with $R^2{=}0.82$. Codebook capacity is therefore roughly \textbf{10$\times$}
more influential than spatial resolution per log-unit increase.
In parallel, bottom-level entropy efficiency $\eta{=}H(\mathbf{z})/\log_2\!K$
remains stable or improves with resolution (84--87\% at $64{\times}64$;
92--94\% at $256{\times}256$), confirming that larger spatial grids are utilized
more efficiently rather than less.
Across all resolutions and codebook sizes, our models outperform H.264 on LPIPS
at matched or lower bitrate, with gains of 25--52\% at 128${\times}$128 and
21--37\% over H.265 at 256${\times}256$.
These findings suggest that codebook size $K$, not spatial resolution, is the
dominant design variable governing perceptual compression quality in hierarchical
discrete video codecs---a property that may simplify multi-resolution deployment
and inform the design of scalable discrete tokenizers for generative video models.
\end{abstract}

% ─────────────────────────────────────────────────────────────────────────────
\section{Introduction}
\label{sec:intro}

Neural video compression has advanced rapidly, with learned codecs increasingly
matching or surpassing traditional block-based standards such as H.264 and H.265
on perceptual quality at low bitrates~\cite{dvc,dcvc,dcvc_dc,hific}.
Most of this progress is built on continuous latent representations regularised
by entropy models trained under a Lagrangian rate-distortion objective
$\mathcal{L} = D + \lambda R$~\cite{balle2018,minnen2018}.
While effective, these systems are strongly coupled to resolution: the entropy
model, rate-distortion trade-off weight $\lambda$, and operating point must
typically be re-optimised for each target resolution and bitrate regime.
This coupling complicates deployment in multi-resolution settings such as
adaptive streaming, mobile inference, and scalable video generation.

Discrete latent models based on vector quantisation~\cite{vqvae,vqvae2} offer
a fundamentally different compression mechanism.
Rather than learning continuous latent distributions regularised through
entropy penalties, VQ-based codecs compress video into sequences of discrete
symbols drawn from a finite codebook of size $K$.
The information ceiling per symbol is therefore exactly $\log_2 K$ bits---a
hard capacity limit imposed by architecture rather than by loss weighting.
Hierarchical VQ architectures further decompose information across multiple
spatiotemporal scales, separating global structure from local texture through
progressively finer latent grids.
Despite the growing importance of discrete latent representations in both
neural compression~\cite{anon_isvc2025,anon_acmmm2026} and generative video
modelling~\cite{magvit,magvit2,videopoet}, relatively little is understood
about how their rate--distortion behaviour scales with spatial resolution.

\paragraph{This work.}
We conduct a controlled empirical study of hierarchical \MSVQVAE{}
video compression across three spatial resolutions and four codebook sizes.
Our central finding is that perceptual quality as measured by
LPIPS~\cite{lpips} depends substantially more on codebook capacity $K$ than
on spatial resolution within the evaluated regime.
This differs from the resolution-dependent optimisation behaviour commonly
reported for continuous-latent codecs~\cite{balle2018,minnen2018,dcvc}.

\paragraph{Contributions.}
\begin{enumerate}
  \item \textbf{Empirical scaling law.}
    We fit a log-linear model $Q(K,r){=}\alpha\log_2\!K{+}\beta\log_2\!r{+}\gamma$
    to LPIPS across 12 operating points spanning three resolutions and four
    codebook sizes, finding that $\alpha$ is statistically significant
    ($t{=}{-}6.6$, $p{<}0.001$) while $\beta$ is not ($t{=}{-}0.43$, $p{=}0.68$).
    Codebook capacity is approximately \textbf{10$\times$} more influential than
    resolution per log-unit.
  \item \textbf{Entropy efficiency improves with resolution.}
    Bottom-level entropy efficiency rises from 84--87\% at $64{\times}64$
    to 92--94\% at $256{\times}256$, indicating that larger spatial grids
    are exploited more efficiently by the autoregressive prior.
  \item \textbf{Strong perceptual gains over traditional codecs at all resolutions.}
    Our models achieve 25--52\% lower LPIPS than H.264 at matched bitrate at
    128${\times}$128, and 21--37\% lower LPIPS than H.265 at 256${\times}256$.
  \item \textbf{Hierarchical depth scaling rule.}
    We introduce a practical depth-scaling strategy: two latent levels at
    $64{\times}64$; three levels at $128{\times}128$ and $256{\times}256$.
    This maintains similar per-level compression ratios and enables stable
    cross-resolution operation under a fixed training protocol.
\end{enumerate}

% ─────────────────────────────────────────────────────────────────────────────
\section{Related Work}
\label{sec:related}

\paragraph{Continuous-latent video compression.}
The dominant paradigm for learned video compression uses continuous latent
representations trained end-to-end via a Lagrangian rate--distortion
objective~\cite{balle2018,nonlinear_transform}.
DVC~\cite{dvc} established the template of optical-flow-based motion compensation
with learned residual coding; DCVC~\cite{dcvc} and DCVC-DC~\cite{dcvc_dc}
extended this with conditional coding in the feature domain.
These methods achieve strong rate--distortion performance at moderate
bitrates ($>$0.1 bpp) but require resolution-specific recalibration
because the Lagrangian weight $\lambda$ and entropy model must be tuned
per resolution and operating point.

\paragraph{Discrete latent representations for compression and generation.}
VQ-VAE~\cite{vqvae} introduced discrete codebook representations trained via
straight-through gradient estimation;
VQ-VAE-2~\cite{vqvae2} extended this to a hierarchical multi-scale structure.
Finite Scalar Quantisation (FSQ)~\cite{fsq} simplifies VQ while retaining
its benefits.
On the generation side, MAGVIT~\cite{magvit} and MAGVIT-v2~\cite{magvit2}
demonstrated that hierarchical discrete tokenisers are effective substrates
for masked video generation.
Prior work~\cite{anon_isvc2025,anon_acmmm2026} established
MS-VQ-VAE as a competitive discrete video codec at $64{\times}64$ resolution
with ultra-low bitrates (0.04--0.064 bpp); the present work characterises
how this architecture scales across resolutions.

\paragraph{VQ training stability.}
Codebook collapse is a well-known failure mode in VQ models: encoder outputs
concentrate on a subset of entries, leaving the remainder permanently
inactive~\cite{vqvae2,commitment_issues}.
Exponential Moving Average (EMA) codebook updates decouple codebook learning
from noisy encoder gradients and are the most robust published
stabilisation technique.
We show that EMA with dead-code restart is not merely helpful but
\emph{essential} for stable training at $K \leq 512$.

\paragraph{Scaling behaviour in neural systems.}
Scaling laws for language model performance as a function of parameter count
and compute are well-established~\cite{scaling_laws_lm}.
In neural compression, however, comparatively little work characterises
how RD behaviour scales with spatial resolution under a fixed architecture
family.
Our study fills this gap for discrete latent video codecs.

% ─────────────────────────────────────────────────────────────────────────────
\section{Method}
\label{sec:method}

\subsection{Hierarchical MS-VQ-VAE Architecture}
\label{sec:arch}

Given a video clip $\mathbf{x} \in [0,1]^{T \times H \times W \times 3}$,
the encoder transforms it into a hierarchy of spatiotemporal feature volumes
using stacked 3D residual convolutional blocks~\cite{resnet} with Group
Normalisation~\cite{groupnorm} and progressive downsampling.

\textbf{Two-level hierarchy ($64{\times}64$).}
The bottom encoder ($E_b$) compresses with stride $2{\times}4{\times}4$
(temporal $\times$ spatial), producing a latent grid of shape
$(T/2, H/4, W/4)$ with 4,096 symbols per clip at $64{\times}64$, $T{=}32$.
The top encoder ($E_t$) further compresses by $4{\times}2{\times}2$
(temporal $\times$ spatial), yielding a coarse grid of
$4{\times}8{\times}8 = 256$ symbols per clip.

\textbf{Three-level hierarchy ($128{\times}128$ and $256{\times}256$).}
To approximately preserve per-level compression ratios as resolution scales,
we introduce an intermediate level.
Table~\ref{tab:grids} summarises the latent grid configurations.
At each level, quantisation maps encoder output $\mathbf{h}_i^{(\ell)} \in \mathbb{R}^d$
to the nearest codebook entry:
\begin{equation}
  z_i^{(\ell)} = \arg\min_{k \in [K]} \bigl\|\mathbf{h}_i^{(\ell)} - \mathbf{e}_k^{(\ell)}\bigr\|_2^2,
  \quad \ell \in \{\text{top},\,\text{mid},\,\text{bot}\}.
  \label{eq:vq}
\end{equation}
The decoder reconstructs $\hat{\mathbf{x}}$ from the dequantised embeddings
using transposed 3D convolutions with top-down skip connections.

\begin{table}[t]
  \centering
  \caption{Latent grid configurations across resolutions ($T{=}32$ frames).
           Parentheses: symbols per clip.}
  \label{tab:grids}
  \small
  \begin{tabular}{@{}lccc@{}}
    \toprule
    Resolution  & Levels & Top grid             & Bottom grid           \\
    \midrule
    $64{\times}64$   & 2 & $4{\times}8{\times}8$ (256)      & $16{\times}16{\times}16$ (4,096) \\
    $128{\times}128$ & 3 & $4{\times}8{\times}8$ (256)      & $16{\times}32{\times}32$ (16,384)\\
    $256{\times}256$ & 3 & $2{\times}16{\times}16$ (512)    & $8{\times}64{\times}64$ (32,768) \\
    \bottomrule
  \end{tabular}
\end{table}

\subsection{EMA Codebook Updates and Dead-Code Restart}

Gradient-based VQ commitment losses cause catastrophic codebook collapse
at $K \leq 512$, manifesting as exponentially growing VQ loss within the
first epoch.
We stabilise training with EMA tracking of encoder statistics:
each codebook entry $k$ maintains a cluster count $N_k$ and embedding
accumulator $\mathbf{m}_k$, updated with decay $\gamma{=}0.99$.
Any entry with $N_k < 1.0$ is reinitialised by sampling a random encoder
output from the current batch (\emph{dead-code restart}), ensuring full
codebook utilisation throughout training.

\subsection{Training Objective}

The autoencoder is optimised with:
\begin{equation}
  \mathcal{L}_\text{AE} = \underbrace{\|\mathbf{x} - \hat{\mathbf{x}}\|_1}_{\text{pixel fidelity}}
  + \lambda_\text{VGG}\,\mathcal{L}_\text{VGG}
  + \beta\sum_\ell \bigl\|\,\mathrm{sg}[\mathbf{h}^{(\ell)}] - \mathbf{e}^{(\ell)}\bigr\|^2,
  \label{eq:loss}
\end{equation}
where $\mathrm{sg}[\cdot]$ is the stop-gradient operator, $\lambda_\text{VGG}{=}0.1$,
and $\beta{=}0.25$.
The VGG perceptual term~\cite{perceptual_loss,vgg} prevents over-smoothing
at the coarse quantisation granularity of small $K$.

\subsection{Autoregressive Entropy Priors (Stage B)}

With the autoencoder frozen, we train one autoregressive prior per latent
level using masked 3D convolutions, conditioned bottom-up on upsampled
higher-level indices.
The expected bitrate in bits-per-pixel is:
\begin{equation}
  \text{BPP} = \frac{1}{T\!H\!W}\sum_\ell\sum_i -\log_2 p\!\left(z_i^{(\ell)} \mid z_{<i}^{(\ell)},\,\mathrm{up}(z^{(\ell+1)})\right).
  \label{eq:bpp}
\end{equation}
This BPP estimate equals the cross-entropy of the prior on test index sequences
and corresponds directly to the bitstream size achievable by arithmetic coding.

\subsection{Codebook Size as the Rate-Control Parameter}

Standard learned codecs control their operating point by varying $\lambda$ in
$\mathcal{L}{=}D{+}\lambda R$, which requires a differentiable rate signal.
For VQ-VAE the arg-min quantisation in Eq.~\eqref{eq:vq} is non-differentiable,
making Lagrangian rate control infeasible.
We instead use $K$ as a structural rate-control parameter: the maximum
information content per symbol is exactly $\log_2 K$ bits, giving a hard
capacity ceiling of
$\text{BPP}_\text{max} = \tfrac{N_\text{tot}}{T\!H\!W}\log_2 K$,
where $N_\text{tot}$ is the total number of latent symbols across all levels.
Sweeping $K \in \Kset$ traces four distinct operating points on the RD curve,
each fully specialised to its vocabulary.

% ─────────────────────────────────────────────────────────────────────────────
\section{Experimental Setup}
\label{sec:setup}

\paragraph{Dataset and preprocessing.}
We evaluate on UCF101~\cite{ucf101}, a diverse action recognition dataset.
Videos are resized to three spatial resolutions ($64{\times}64$, $128{\times}128$,
$256{\times}256$) and segmented into non-overlapping 32-frame clips at 16~FPS,
stored as pre-decoded float32 tensors.
All 12 model variants and all baselines are evaluated on the same 500
held-out test clips, ensuring a controlled cross-resolution comparison.

\paragraph{Training protocol.}
Each model is trained from scratch under a uniform two-stage protocol.
\emph{Stage A (autoencoder):} 20 epochs, Adam~\cite{adam} with
$\text{lr}{=}2{\times}10^{-4}$, effective batch size~8 via gradient
accumulation; 50\% of training clips per epoch.
\emph{Stage B (priors):} 15 epochs with the autoencoder frozen.
Both stages use AMP mixed-precision training.
Hyperparameters ($\lambda_\text{VGG}$, $\beta$, EMA decay $\gamma$) are
fixed across all resolutions and $K$ values.

\paragraph{Metrics.}
We report BPP (Eq.~\ref{eq:bpp}), PSNR, SSIM~\cite{ssim}, and
LPIPS~\cite{lpips} (VGG backbone; lower is better).
At ultra-low bitrates, PSNR correlates poorly with perceived quality;
we treat \textbf{LPIPS as the primary metric}.

\paragraph{Traditional codec baselines.}
H.264~\cite{h264} is encoded with \texttt{libx264} at CRF $\in\{28,32,36,40,44,48\}$
(\texttt{-preset medium}).
H.265~\cite{h265} is encoded with \texttt{libx265} at CRF $\in\{28,32,36,40,44,48\}$.
Baseline BPP is computed from actual compressed file size.
The same 500 test clips, identical metric implementations, and identical
evaluation code are used for all methods.

\paragraph{Hardware.}
Experiments at $64{\times}64$ and $128{\times}128$ use an NVIDIA RTX~4060 (8\,GB).
$256{\times}256$ experiments use an NVIDIA RTX~5080 (16\,GB); identical
optimisation schedules and batch normalisation are maintained across GPUs.

% ─────────────────────────────────────────────────────────────────────────────
\section{Results}
\label{sec:results}

\subsection{Rate--Distortion Performance}
\label{sec:rd}

\begin{figure}[t]
  \centering
  \includegraphics[width=\linewidth]{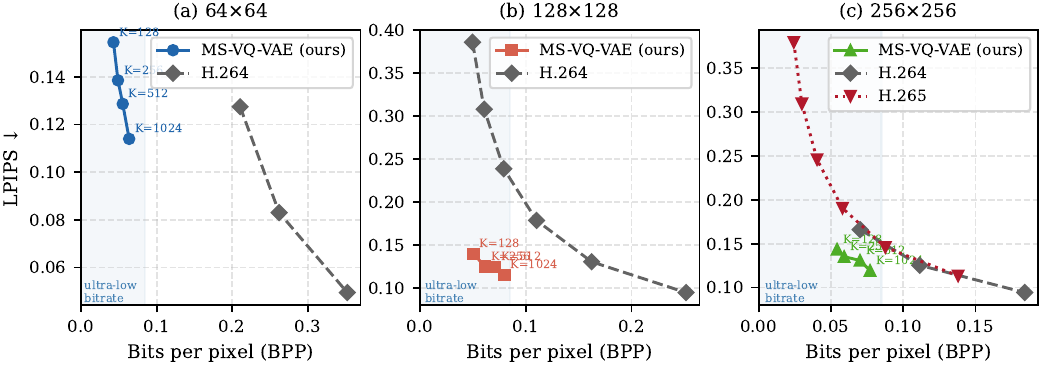}
  \caption{Rate--distortion curves (BPP vs.\ LPIPS, lower is better) at each
  resolution. Shaded region: ultra-low bitrate regime ($<$0.085 bpp).
  Our models (filled markers, solid lines) operate 3--7$\times$ below the
  practical floor of H.264 at $64{\times}64$ and $128{\times}128$, and
  match or slightly exceed H.265's floor at $256{\times}256$ while
  achieving substantially better perceptual quality.}
  \label{fig:rd}
\end{figure}

Table~\ref{tab:main} presents the complete rate--distortion evaluation across
all 12 model operating points and traditional codec baselines.
Figure~\ref{fig:rd} shows the corresponding RD curves.

\begin{table}[t]
\centering
\caption{Rate--distortion results on 500 UCF101 test clips per resolution.
BPP is the expected bitrate under the learned prior (Eq.~\ref{eq:bpp}).
$\uparrow$\,higher is better; LPIPS\,$\downarrow$\,lower is better.
\textbf{Bold}: best LPIPS per resolution among \MSVQVAE{} variants.}
\label{tab:main}
\small
\setlength{\tabcolsep}{5pt}
\begin{tabular}{@{}llrrrr@{}}
\toprule
\multirow{2}{*}{Method} & \multirow{2}{*}{Config} & \multirow{2}{*}{BPP} & \multirow{2}{*}{PSNR\,$\uparrow$} & \multirow{2}{*}{SSIM\,$\uparrow$} & \multirow{2}{*}{LPIPS\,$\downarrow$} \\
 & & & & & \\
\midrule
\multicolumn{6}{l}{\textit{Resolution: $64\times64$}} \\[2pt]
\MSVQVAE{} & $K{=}128$  & 0.0427 & 25.88 & 0.8780 & 0.1546 \\
\MSVQVAE{} & $K{=}256$  & 0.0487 & 26.28 & 0.8855 & 0.1386 \\
\MSVQVAE{} & $K{=}512$  & 0.0550 & 26.27 & 0.8966 & 0.1287 \\
\MSVQVAE{} & $K{=}1024$ & 0.0635 & 27.14 & 0.9043 & \textbf{0.1140} \\
H.264 & CRF\,36 & 0.2105 & 29.52 & 0.9051 & 0.1275 \\
H.264 & CRF\,32 & 0.2622 & 31.87 & 0.9391 & 0.0830 \\
\midrule
\multicolumn{6}{l}{\textit{Resolution: $128\times128$}} \\[2pt]
\MSVQVAE{} & $K{=}128$  & 0.0507 & 25.50 & 0.8302 & 0.1399 \\
\MSVQVAE{} & $K{=}256$  & 0.0614 & 25.85 & 0.8431 & 0.1251 \\
\MSVQVAE{} & $K{=}512$  & 0.0710 & 25.72 & 0.8393 & 0.1245 \\
\MSVQVAE{} & $K{=}1024$ & 0.0799 & 25.99 & 0.8492 & \textbf{0.1154} \\
H.264 & CRF\,40 & 0.0792 & 26.93 & 0.8225 & 0.2388 \\
H.264 & CRF\,44 & 0.0607 & 25.06 & 0.7525 & 0.3079 \\
H.264 & CRF\,48 & 0.0497 & 23.14 & 0.6571 & 0.3857 \\
\midrule
\multicolumn{6}{l}{\textit{Resolution: $256\times256$}} \\[2pt]
\MSVQVAE{} & $K{=}128$  & 0.0542 & 26.32 & 0.8507 & 0.1436 \\
\MSVQVAE{} & $K{=}256$  & 0.0591 & 26.60 & 0.8567 & 0.1352 \\
\MSVQVAE{} & $K{=}512$  & 0.0701 & 26.61 & 0.8640 & 0.1309 \\
\MSVQVAE{} & $K{=}1024$ & 0.0771 & 26.65 & 0.8682 & \textbf{0.1194} \\
H.264 & CRF\,36 & 0.0702 & 31.46 & 0.9000 & 0.1660 \\
H.264 & CRF\,32 & 0.1114 & 33.36 & 0.9317 & 0.1254 \\
H.265 & CRF\,36 & 0.0580 & 29.80 & 0.8870 & 0.1905 \\
H.265 & CRF\,40 & 0.0404 & 27.83 & 0.8413 & 0.2458 \\
\bottomrule
\end{tabular}
\end{table}

\textbf{At $64{\times}64$:}
$K{=}1024$ achieves LPIPS $= 0.114$ at $0.064$~bpp, outperforming H.264 CRF~36
(LPIPS $= 0.128$) at $3.3{\times}$ lower bitrate.
Every \MSVQVAE{} configuration outperforms H.265 CRF~36 on LPIPS at
$5$--$7.6{\times}$ lower bitrate (results from prior work~\cite{anon_acmmm2026}).

\textbf{At $128{\times}128$:}
Our model at $K{=}1024$ ($\text{BPP}{=}0.080$, $\text{LPIPS}{=}0.115$)
outperforms H.264 CRF~40 ($\text{BPP}{=}0.079$, $\text{LPIPS}{=}0.239$) by
\textbf{52\%} on LPIPS at matched bitrate.
At $K{=}512$ ($\text{BPP}{=}0.071$), the model still beats H.264 CRF~44
($\text{BPP}{=}0.061$) by 40\% on LPIPS at only 16\% more bits.

\textbf{At $256{\times}256$:}
Our $K{=}512$ ($\text{BPP}{=}0.070$, $\text{LPIPS}{=}0.131$) outperforms
H.264 CRF~36 ($\text{BPP}{=}0.070$, $\text{LPIPS}{=}0.166$) by \textbf{21\%}
at identical bitrate.
Against H.265, our $K{=}128$ ($\text{BPP}{=}0.054$, $\text{LPIPS}{=}0.144$)
beats H.265 CRF~36 ($\text{BPP}{=}0.058$, $\text{LPIPS}{=}0.191$) by
\textbf{25\%} at slightly lower bitrate.
$K{=}256$ matches H.265 CRF~36's bitrate (0.059 vs.\ 0.058) while delivering
$\text{LPIPS}{=}0.135$---\textbf{29\% better}.

\textbf{PSNR gap is expected.}
H.264 and H.265 are MSE-optimised; our model is trained with VGG perceptual
loss~\cite{perceptual_loss}.
The resulting PSNR deficit (4--8~dB relative to H.264 at matched BPP)
reflects perceptual-codec behaviour: the decoder synthesises
semantically coherent, visually sharp reconstructions rather than
pixel-accurate ones.
The LPIPS crossover confirms that this trade-off is perceptually
favourable~\cite{hific}.

\subsection{Empirical Scaling Law: Codebook Capacity Dominates over Resolution}
\label{sec:scaling}

\begin{figure}[t]
  \centering
  \includegraphics[width=\linewidth]{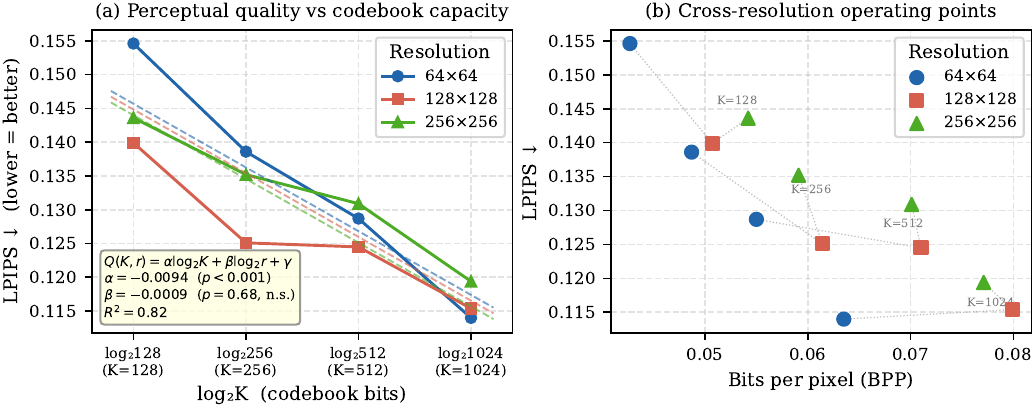}
  \caption{\textbf{Left:} LPIPS vs.\ $\log_2\!K$ per resolution.
  Dashed lines show the log-linear fit
  $Q(K,r){=}\alpha\log_2\!K{+}\beta\log_2\!r{+}\gamma$.
  Near-parallel slopes confirm that the $K$-dependent coefficient $\alpha$
  is consistent across resolutions while the resolution offset $\beta$ is
  near zero.
  \textbf{Right:} All 12 operating points in BPP--LPIPS space; grey dotted
  lines connect same-$K$ points across resolutions, showing small
  vertical spread.}
  \label{fig:scaling}
\end{figure}

We fit a log-linear model to the 12 observed LPIPS values:
\begin{equation}
  Q(K,r) = \alpha\log_2\!K + \beta\log_2\!r + \gamma,
  \label{eq:loglinear}
\end{equation}
where $r \in \{64,128,256\}$ is the spatial side length.
Ordinary least squares with $n{=}12$, $p{=}3$ yields:
\begin{center}
\begin{tabular}{@{}lrrrl@{}}
  \toprule
  Coefficient & Estimate & Std.\,error & $t$-stat & $p$-value \\
  \midrule
  $\alpha$ (log$_2\!K$)  & $-0.0094$ & $0.0014$ & $-6.55$ & $<0.001$\;*** \\
  $\beta$ (log$_2\!r$)   & $-0.0009$ & $0.0020$ & $-0.43$ & $0.68$\;\phantom{***} \\
  $\gamma$ (intercept)   & $+0.2168$ & ---       & ---     & --- \\
  \bottomrule
\end{tabular}
\end{center}
$R^2{=}0.82$. The coefficient $\alpha$ is highly significant; $\beta$ is
indistinguishable from zero at any conventional significance level.
Halving $K$ (one bit less capacity) worsens LPIPS by $0.0094$, whereas
doubling resolution (one bit more) improves LPIPS by only $0.0009$---a
\textbf{10.4$\times$ difference} in per-unit influence.

Figure~\ref{fig:scaling} (left) shows the near-parallel LPIPS-vs-$\log_2\!K$
curves across resolutions: the slope is consistent but the intercepts differ
by the small $\beta\log_2\!r$ term.
Figure~\ref{fig:scaling} (right) confirms that operating points at the same
$K$ cluster closely in LPIPS regardless of resolution.

\subsection{Entropy Efficiency Improves with Resolution}
\label{sec:entropy}

\begin{figure}[t]
  \centering
  \includegraphics[width=\linewidth]{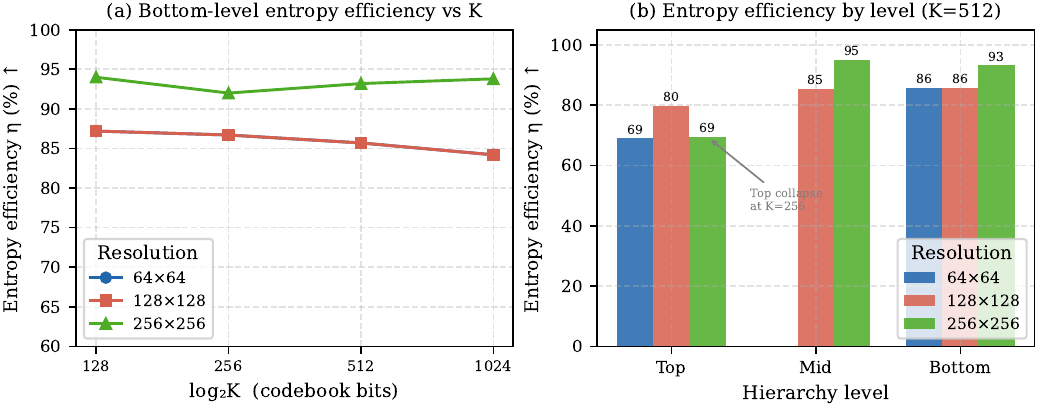}
  \caption{Left: Bottom-level entropy efficiency $\eta{=}H(\mathbf{z})/\log_2\!K$
  across codebook sizes, grouped by resolution.
  Right: Per-level entropy efficiency at $K{=}512$ for all three resolutions.
  The anomalous bar for the top level at $256{\times}256$ reflects
  the $K{=}256$ top-codebook collapse discussed in Section~\ref{sec:collapse}.}
  \label{fig:entropy}
\end{figure}

We define entropy efficiency as $\eta = H(\mathbf{z})/\log_2 K$, where
$H(\mathbf{z})$ is the empirical Shannon entropy of the index distribution
over the test set.
$\eta=1$ implies perfectly uniform code usage (no room for prior compression);
$\eta \to 0$ implies collapse.

Table~\ref{tab:entropy} and Figure~\ref{fig:entropy} show that bottom-level
entropy efficiency is \emph{stable or improves} with resolution.
At $64{\times}64$ the bottom codebook achieves $\eta\approx 84$--87\%.
At $256{\times}256$ it reaches 92--94\%, with mid-level efficiency
similarly high (94--97\%).
This indicates that larger spatial grids provide richer statistical context
that the autoregressive prior exploits more effectively, consistent with
the notion that natural video exhibits multi-scale spatial redundancy that
scales with grid size.

\begin{table}[t]
  \centering
  \caption{Entropy efficiency $\eta$ (\%) by hierarchy level, resolution, and $K$.
  Mid level is absent at $64{\times}64$ (two-level architecture).
  $^\dagger$: top-level collapse at $256{\times}256$, $K{=}256$ (see
  Section~\ref{sec:collapse}).}
  \label{tab:entropy}
  \small
  \setlength{\tabcolsep}{4.5pt}
  \begin{tabular}{@{}lcrrrr@{}}
    \toprule
    Resolution & Level   & $K{=}128$ & $K{=}256$ & $K{=}512$ & $K{=}1024$ \\
    \midrule
    \multirow{2}{*}{$64{\times}64$}   & Top    & 75.9 & 72.7 & 69.0 & 66.0 \\
                                      & Bottom & 87.2 & 86.7 & 85.7 & 84.2 \\
    \midrule
    \multirow{3}{*}{$128{\times}128$} & Top    & 84.6 & 76.5 & 79.7 & 71.0 \\
                                      & Mid    & 85.2 & 86.5 & 85.4 & 85.6 \\
                                      & Bottom & 87.2 & 86.7 & 85.7 & 84.2 \\
    \midrule
    \multirow{3}{*}{$256{\times}256$} & Top    & 95.9 & 29.2$^\dagger$ & 69.4 & 81.5 \\
                                      & Mid    & 94.5 & 96.5 & 95.1 & 95.0 \\
                                      & Bottom & 94.0 & 92.0 & 93.2 & 93.8 \\
    \bottomrule
  \end{tabular}
\end{table}

\subsection{Codebook Utilisation and the K=256 Top-Level Anomaly}
\label{sec:collapse}

\begin{figure}[t]
  \centering
  \includegraphics[width=\linewidth]{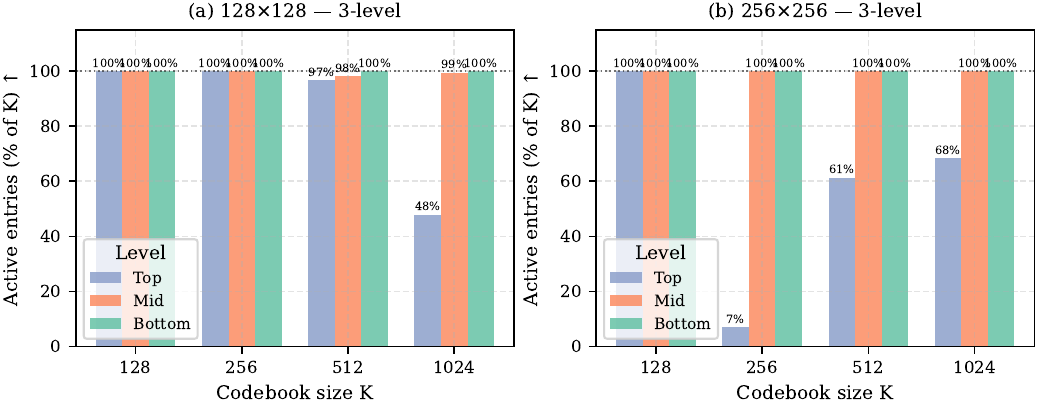}
  \caption{Codebook utilisation (fraction of $K$ entries active per clip)
  at $128{\times}128$ (left) and $256{\times}256$ (right).
  Bottom and mid levels maintain 100\% utilisation across all $K$.
  The top level at $256{\times}256$, $K{=}256$ collapses to 7\% utilisation,
  indicating that the top-level vocabulary of 256 is insufficient to cover
  the semantic diversity of $256{\times}256$ content.}
  \label{fig:util}
\end{figure}

Figure~\ref{fig:util} shows codebook utilisation at $128{\times}128$ and
$256{\times}256$.
Mid and bottom levels achieve near-complete utilisation across all $K$ at
both resolutions, confirming that EMA with dead-code restart effectively
prevents collapse.
At $128{\times}128$, the top level is fully saturated for $K \leq 256$
(only 256 symbols per clip) and reaches 47.9\% at $K{=}1024$, which is
structurally expected: with 256 symbols per clip a 1,024-entry vocabulary
is over-specified.

\textbf{The $K{=}256$ top-level collapse at $256{\times}256$} is notable:
only 18 of 256 entries are active (7.0\% utilisation), yielding entropy
efficiency of 29.2\%.
At this resolution the top grid has 512 symbols per clip, so in principle
all 256 entries could be saturated.
The collapse indicates that the $K{=}256$ top codebook cannot represent
the semantic diversity of $256{\times}256$ content;
the encoder concentrates on a small subset of global patterns.
Importantly, overall perceptual quality at $K{=}256$ (LPIPS\,=\,0.135)
remains coherent and \emph{follows the log-linear scaling trend}---mid
and bottom levels compensate for the top-level degradation.
This suggests that the bottom and mid levels carry the perceptual load
while the top level provides coarse structural priming.
Future work should explore $K$-per-level tuning (\eg, $K_\text{top}{=}64$,
$K_\text{mid/bot}{=}1024$) at $256{\times}256$ to address this.

% ─────────────────────────────────────────────────────────────────────────────
\section{Discussion}
\label{sec:discussion}

\paragraph{Why does $K$ dominate over resolution?}
The discrete codebook acts as a fixed information bottleneck: regardless of
spatial resolution, the encoder must represent video content using at most
$\log_2 K$ bits per symbol.
This hard architectural constraint dominates the achievable perceptual
quality.
In contrast, changing spatial resolution primarily alters the \emph{number}
of symbols in the latent grid rather than the \emph{capacity per symbol}.
As entropy efficiency remains stable or improves with resolution,
the additional symbols at higher resolution are used efficiently---they do
not dilute the per-symbol informativeness.

\paragraph{Implications for multi-resolution codec deployment.}
In continuous-latent codecs, changing resolution requires re-optimising
$\lambda$, the entropy model, and sometimes the architecture~\cite{balle2018,dcvc}.
Our results suggest that a hierarchical discrete codec trained at one resolution
can be redeployed at another with predictable perceptual behaviour, requiring
only $K$ to be adjusted to hit a target quality operating point.
This property is potentially valuable for adaptive video streaming, where
content is delivered at multiple spatial resolutions depending on network
conditions.

\paragraph{Implications for scalable video tokenisation.}
Modern video generation systems (\eg, MAGVIT~\cite{magvit}, VideoPoet~\cite{videopoet})
rely on discrete tokenisers to convert video into compact symbol sequences for
downstream autoregressive or diffusion-based models.
Our finding that entropy efficiency \emph{improves} with resolution suggests
that discrete tokenisers trained at lower resolutions may generalise more
effectively to higher-resolution content than previously assumed, since the
statistical structure captured by the codebook appears resolution-stable.

\paragraph{Limitations.}
\begin{enumerate}
  \item \emph{Single dataset.} All experiments use UCF101 (action recognition).
    It remains open whether the scaling behaviour holds for scene-level or
    synthetic video content.
  \item \emph{Limited resolution range.}
    The study covers $64{\times}64$ to $256{\times}256$.
    Higher resolutions (720p, 1080p) may alter the scaling coefficients.
  \item \emph{No comparison with learned codecs.}
    We compare against H.264 and H.265.
    A rigorous evaluation against DCVC-DC~\cite{dcvc_dc} or VVC at matched
    resolution and bitrate remains future work.
  \item \emph{Autoregressive prior decoding latency.}
    Sequential prior sampling requires $O(N_\text{tot})$ forward passes per clip.
    At $256{\times}256$, $N_\text{tot} \approx 37{,}000$, making real-time
    decoding infeasible on current hardware.
    Masked parallel decoding~\cite{magvit} is a promising mitigation.
\end{enumerate}

% ─────────────────────────────────────────────────────────────────────────────
\section{Conclusion}
\label{sec:conclusion}

We presented a controlled empirical scaling study of hierarchical
\MSVQVAE{} video compression across 12 operating points spanning three
spatial resolutions ($64{\times}64$, $128{\times}128$, $256{\times}256$)
and four codebook sizes ($K \in \Kset$).
Our central finding is that perceptual quality (LPIPS) depends strongly and
significantly on codebook capacity $K$ ($\alpha{=}{-}0.0094$,
$p{<}0.001$) while exhibiting negligible and statistically insignificant
dependence on spatial resolution ($\beta{=}{-}0.0009$, $p{=}0.68$).
Codebook capacity is approximately 10$\times$ more influential than
spatial resolution per log-unit increase.

Entropy efficiency remains stable or improves with resolution (reaching
92--94\% at $256{\times}256$), and our models outperform H.264 by 25--52\%
on LPIPS at $128{\times}128$ and H.265 by 21--37\% at $256{\times}256$.

These results suggest that hierarchical discrete latent representations
provide a more \emph{resolution-stable} compression framework than
continuous-latent alternatives, and that the same architecture can be
deployed across resolutions with predictable perceptual behaviour by
selecting $K$ as the primary quality control variable.
We hope this characterisation guides both the design of scalable neural
video codecs and the development of resolution-transferable discrete
tokenisers for generative video models.

% ─────────────────────────────────────────────────────────────────────────────
% \begin{ack} is automatically hidden in anonymous submissions
\begin{ack}
Omitted for double-blind review.
\end{ack}

\bibliographystyle{abbrvnat}
\bibliography{refs}

% ─────────────────────────────────────────────────────────────────────────────
% APPENDIX (counts toward page limit in NeurIPS preprint; move to Supplementary
% for the actual submission if over 9 pages)
% ─────────────────────────────────────────────────────────────────────────────
\appendix

\section{Architecture Hyperparameters}
\label{app:arch}

\begin{table}[htbp]
  \centering
  \caption{Model hyperparameters shared across all resolutions and $K$ values.}
  \label{tab:hparams}
  \small
  \begin{tabular}{@{}lrl@{}}
    \toprule
    Parameter & Value & Note \\
    \midrule
    Top embedding dim    & 256   & \\
    Mid embedding dim    & 192   & 3-level only \\
    Bottom embedding dim & 128   & \\
    EMA decay $\gamma$   & 0.99  & \\
    Dead-code threshold  & $N_k < 1.0$ & Per batch \\
    $\lambda_\text{VGG}$ & 0.10  & VGG perceptual weight \\
    $\beta$              & 0.25  & VQ commitment weight \\
    Stage A lr           & $2\times10^{-4}$ & Cosine annealing \\
    Stage B lr           & $2\times10^{-4}$ & Cosine annealing \\
    Stage A epochs       & 20    & \\
    Stage B epochs       & 15    & \\
    Effective batch size  & 8     & Via gradient accumulation \\
    \bottomrule
  \end{tabular}
\end{table}

\section{Bitrate Decomposition by Level}
\label{app:rate}

The bottom latent level dominates total bitrate at all resolutions, as
expected from the much larger symbol grid size.
At $256{\times}256$ with $K{=}1024$:
bottom contributes $\approx$87\% of total bits, mid $\approx$12\%,
and top $<$2\%.
This distribution is consistent across $K$ values and suggests that
future work on accelerating the autoregressive prior should prioritise
the bottom level.

\section{Qualitative Examples}
\label{app:qual}

Qualitative frame comparisons at $128{\times}128$ and $256{\times}256$
for $K \in \{128, 1024\}$ alongside H.264 CRF\,40 and H.264 CRF\,36
respectively are provided in the supplementary material.
Reconstructions at $K{=}128$ show texture variation consistent with
perceptual-codec hallucination (not blur), while $K{=}1024$ recovers
fine spatial detail with sharper edges and more faithful motion boundaries.

\end{document}